\documentclass[aps,twocolumn,
groupedaddress
,longbibliography,prl]{revtex4-2}

\usepackage{amsmath,amssymb}
\usepackage{graphicx}
\usepackage{multirow}
\usepackage{bm}
\usepackage{mathtools}
\usepackage{dsfont}
\usepackage{amsfonts}
\usepackage{xcolor}
\usepackage[normalem]{ulem}
\usepackage{url}
\usepackage{wasysym}
\usepackage{bbm}

\usepackage[colorlinks=true,linkcolor=magenta,citecolor=magenta,hypertexnames=false]{hyperref}
\newcommand{\nocontentsline}[3]{}
\newcommand{\tocless}[2]{\bgroup\let\addcontentsline=\nocontentsline#1{#2}\egroup}

\def\ba#1\ea{\begin{align}#1\end{align}}
\def\bg#1\eg{\begin{gather}#1\end{gather}}
\def\bpm{\begin{pmatrix}}
\def\epm{\end{pmatrix}}

\newcommand{\sgn}{{\rm sgn}}

\newcommand{\ket}[1]{|#1\rangle}
\newcommand{\bra}[1]{\langle#1|}

\newcommand{\magenta}[1]{\textcolor{magenta}{#1}}

\newcommand{\ourtitle}{
Obstruction to Broken Symmetries in Topological Flat Bands}

\allowdisplaybreaks

\begin{document}
\title{\textbf{\ourtitle}}

\author{Penghao Zhu}
\author{Shi Feng}
\author{Yuan-Ming Lu}
\affiliation{Department of Physics, The Ohio State University, Columbus, OH 43210, USA}


\begin{abstract}
Motivated by the abundance of symmetry breaking states in magic-angle twisted bilayer graphene and other two-dimensional materials, we study superconducting (SC) and charge orders in two-dimensional topological flat bands in the strong correlation regime. By relating the half-filled 2D topological flat bands to the surface states of 3D topological insulators in symmetry class AIII, we reveal the topological obstruction to the formation of gapped SC and inter-valley charge orders without intrinsic topological orders, in the presence of the anti-unitary particle-hole symmetry at half filling. This is a generalization of the Li-Haldane arguments for nodal superconductivity to strongly interacting electrons. In contrast to the $\mathbb{Z}$-valued obstruction derived from the non-interacting band topology, the topological obstruction of interacting electrons in half-filled flat bands has a $\mathbb{Z}_{8}$ classification, depending on the charge (valley) Chern number of the superconducting (inter-valley charge) orders. This is demonstrated by an interacting Hamiltonian for half-filled flat bands with a net Chern number $C=4$, where superconductivity and $\mathbb{Z}_2$ topological order coexist in a gapped ground state with particle-hole symmetry.
\end{abstract}

\maketitle

\let\oldaddcontentsline\addcontentsline
\renewcommand{\addcontentsline}[3]{}
\magenta{\it Introduction.|}  
Topological phases of matter are characterized by their robust properties stemming from the quantum entanglement of their ground states~\cite{Zeng2019B}. While topological phases cannot be distinguished and classified by local order parameters resulting from spontaneous symmetry breaking, there is a close relationship between topological phases and long-range orders. The presence of long-range orders facilitates the emergence of topological phases by establishing the topological gaps, as can be seen from the following examples: (i) Magnetic orders are crucial for quantum anomalous hall phases~\cite{haldane1988model,chang2013experimental,deng2020quantum}; (ii) Superconducting orders are pertinent to topological superconductors hosting Majorana zero modes, where the Bogoliubov-de Gennes (BdG) quasiparticles exhibit nontrivial topology~\cite{schnyder2008classification,kitaev2009periodic, qi2011topological, Sato2017topological}; (iii) Gaps associated with charge orders, such as Peierls charge density waves in 1D, can result in topological insulators with quantized polarizations~\cite{su1979solitons,asboth2016short}; (iv) Topological Mott insulators spontaneously break the spin rotational symmetry~\cite{raghu2008topological}, facilitating the spin-orbit couplings needed for topological insulators\cite{Hasan2010,Hasan2011,Qi2011}. With compelling evidence demonstrating the one-way influence of long-range orders on topological phases, it is natural to ask the following fundamental question: How can the topology affect the long-range order in the ground state?

One well-known example of such is the BCS-type superconductivity (SC) in a magnetic Weyl semimetal, where pairing order parameter between two Fermi surfaces (FSs) enclosing two Weyl points of opposite chirality must be nodal~\cite{murakami2003berry,Li2018topological}. Interestingly, there is a topological obstruction to a gapped/nodeless SC state due to the Chern number of paired electrons on the FSs. Inspired by this result, topological obstruction to gapped superconductors have been discussed for pairings between electrons in normal states with nontrivial $\mathbb{Z}_2$ topology~\cite{Sun2020mathbb}, fragile topology~\cite{Yu2022euler,yu2023euler}, and delicate topology~\cite{zhu2024dipole}. Moreover, topologically obstructed gapless/nodal charge order has also been studied for electrons in topological bands with nonzero Chern numbers~\cite{bobrow2020monopole,bultinck2020mechanism}. The fundamental idea behind these works is to establish a connection between the vorticity of the order parameters and the topological invariants of electronic bands. As a result, the presence of nontrivial topological invariants indicates the existence of nonvanishing vorticities of the order parameters in momentum space, and thus zero points of the order parameters. These arguments rely on the mean-field picture by analyzing the geometry and topology of non-interacting Bloch wavefunctions, which breaks down in the presence of strong interactions. Therefore, a natural question is: do the topological obstruction to gapped SC and charge orders persist in strongly correlated systems? Besides bearing the theoretical importance, addressing this question is also helpful for understanding ordered phases in realistic materials, given that the topologically obstructed SC order and inter-valley coherent (IVC) order have been proposed to occur in flat bands of Moire systems based on band theory~\cite{yu2023euler,wang2024molecular,bultinck2020mechanism}, where the interactions are strong compared to the small band width. 

In this work, we study the SC/charge orders arising from half-filled electrons in 2D flat bands with nonzero charge/valley Chern numbers that will be defined later. By mapping these electronic states to the surface states of 3D interacting symmetry protected topological (SPT) phases, we show that the SC order and charge order cannot be symmetrically gapped out without developing topological orders due to the bulk topology of the 3D SPTs. This topological obstruction to gapped long-range orders can be viewed as a generalization of the Lieb-Schultz-Mattis type theorems~\cite{lieb1961two,oshikawa2000commensurability} to long-range ordered phases. Furthermore, we demonstrate that these topologically obstructed long-range orders have a $\mathbb{Z}_{8}$ classification, in contrast with the $\mathbb{Z}$ classification derived from the non-interacting band theory. To demonstrate the topological obstruction in correlated electrons, we construct an interacting Hamiltonian for a set of flat bands with net Chern number $C=4$, and numerically obtain a symmetrically gapped ground state, where SC long-range order and $Z_2$ topological order coexist. 







\begin{figure}[h]
\centering
\includegraphics[width=0.9\columnwidth]{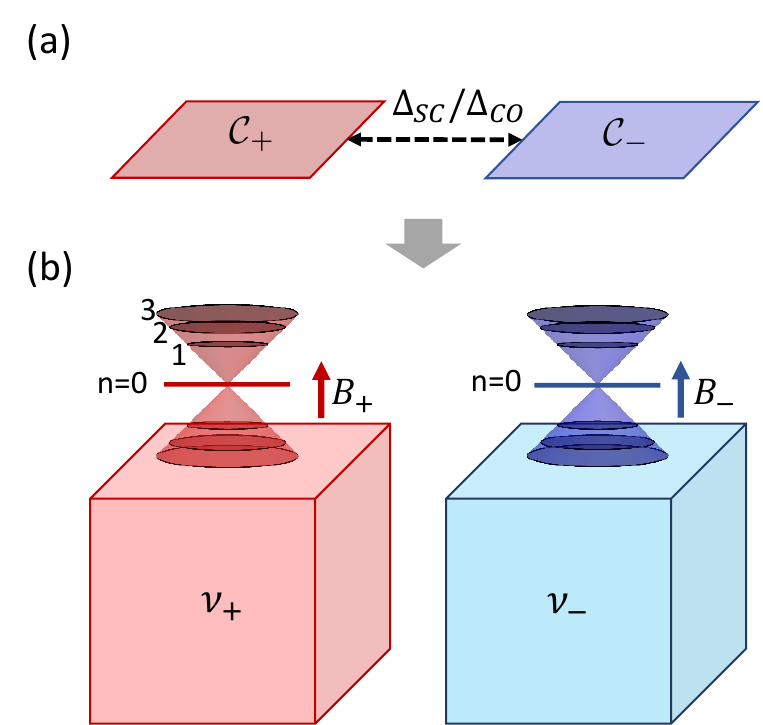}
\caption{(a)Illustration inter-valley SC and charge order for two flat Chern bands with Chern number $C_{\pm}$. Focusing on the $|C_{\pm}|=1$ case, the two flat Chern bands can be mapped onto the zeroth LLs of the surface Dirac modes of 3D SPTs with bulk topological index $\nu_{\pm}=C_\pm$ shown in (b). The $0,1,2,3$ in (b) labels different LLs. }
\label{fig:illustration}
\end{figure}

\magenta{\it Problem setup and symmetry.|} As illustrated in Fig.~\ref{fig:illustration}(a), we start by considering two degenerate flat Chern bands (two ``valleys''), $c^{\dag}_{+,\mathbf{k}}$ and $c^{\dag}_{-,\mathbf{k}}$, labelled by valley index $+$ and $-$.  The Chern numbers of these two bands are respectively $\mathcal{C}_{+}$ and $\mathcal{C}_{-}$. We study the topological obstructions for \textit{inter-valley} SC and charge orders in these half-filled flat bands, of which the order parameters are $\Delta_{\rm SC}(\mathbf{k})=\langle c_{+,\mathbf{k}}c_{-,-\mathbf{k}} \rangle$ and $\Delta_{\rm CO}(\mathbf{k})=\langle c^{\dag}_{+,\mathbf{k}}c_{-,\mathbf{k}} \rangle$, respectively.


At half filling, by projecting the Hilbert space onto these two flat Chern bands, there will be an emergent, anti-unitary particle-hole (i.e. chiral) symmetry ($\hat{S}$), of which the representation is chosen to be
\begin{equation}
\label{eq:chiral}
\hat{S}c^{\dag}_{\pm,\mathbf{k}}\hat{S}^{-1}=\pm c_{\pm,\mathbf{k}},\ \hat{S}i\hat{S}^{-1}=-i.
\end{equation}
The above choice of chiral symmetry $\hat S$ forbids intra-valley terms such as $\tilde{\Delta}_{1}c_{+,\mathbf{k}}c_{+,-\mathbf{k}}+h.c.$ and $\tilde{\Delta}_{2}c^{\dag}_{+,\mathbf{k}}c_{+,\mathbf{k}}+h.c.$, and automatically selects the inter-valley orders. Note that although the chiral symmetry seems to impose strong constraints on interacting Hamiltonians, it is actually preserved for generic two-body interactions (up to a one-body term) at half filling and thus is relevant for realistic systems~\cite{wang2016half}. A detailed discussion of this point can be found in Supplementary Material (SM)~\cite{supp}. Besides the chiral symmetry, we also consider the charge $U(1)$ and the valley $U(1)$ symmetries, denoted as $U_c(1)$ and $U_{v}(1)$, which respectively correspond to the conservation of the net charge and the charge difference between the $\pm$ valleys. Importantly, the SC and charge orders correspond to spontaneous breaking of $U_c(1)$ and $U_{v}(1)$, respectively. With these symmetries in mind, we can now concretely formulate the topological obstructions to inter-valley SC and charge orders as the following question:
\begin{itemize}
    \item Starting with a half-filled interacting electron system with $U_{c}(1)\times U_{v}(1)\times Z_2^{\hat{S}}$ symmetries, after breaking either $U_c(1)$ or $U_v(1)$ symmetry, is there a gapped ground state of SC or charge order that preserves the remaining symmetries?
\end{itemize}


Applying the non-interacting band theory at the mean-field level, it can be proven that the net vorticity of $\Delta_{\rm SC}(\mathbf{k})$ ($\Delta_{\rm CO}(\mathbf{k})$), which is defined as $vort=(1/2\pi)\sum_{\text{vortex}}\oint\partial_{\mathbf{k}}\operatorname{arg}\Delta_{\rm SC(CO)}(\mathbf{k})$, is determined by the Chern numbers of the two sectors through $vort=\mathcal{C}_{+}+\mathcal{C}_{-}$ ($vort=\mathcal{C}_{+}-\mathcal{C}_{-}$)~\cite{murakami2003berry,Li2018topological,Yu2022euler,yu2023euler,bobrow2020monopole,bultinck2020mechanism,zhu2024dipole}. A review of this Chern-vorticity theorem can be found in the SM~\cite{supp}. 
When $vort$ is nonzero, there must be zero points of $\Delta_{\rm SC}(\mathbf{k})$ ($\Delta_{\rm CO}(\mathbf{k})$) in the momentum space. Consequently, the corresponding quadratic term $\Delta_{\rm SC}(\mathbf{k})c^{\dag}_{+,\mathbf{k}}c^{\dag}_{-,-\mathbf{k}}+{\rm H.c.}$ ($\Delta_{\rm CO}(\mathbf{k})c^{\dag}_{+,\mathbf{k}}c_{-,\mathbf{k}}+{\rm H.c.}$), which is the only chiral symmetric SC pairing (charge order) term, cannot give rise to a full gap in the fermion spectrum. Equivalently, this means that there is no chiral symmetric, gapped ground state with SC or charge order. Note that the above Chern-vorticity theorem points to an integer($\mathbb{Z}$)-valued  obstruction to gapped SC/charge order, since the net vorticity is $\mathbb{Z}$-valued.

\magenta{\it Topological obstructions in strongly correlated systems.|} 
In the strong correlation regime, however, the above proof for Chern-vorticity theorem based on the band theory breaks down, calling for a new approach to establish topological obstruction in interacting systems. To this end, we first focus on the simplest case where $|C_{\pm}|=1$, and map the flat Chern bands onto the surface states of 3D class AIII fermionic topological insulators (i.e. 3D fermionic SPTs with $U(1)\times Z_2^{\hat{S}}$ symmetry) following the idea introduced in Ref.~\cite{wang2016half,potter2017real}. Through this map, we can associate the topological obstruction for gapped SC/charge-ordered ground states to the bulk topology of the 3D SPT, because a nontrivial 3D bulk guarantees the surface, if preserving the bulk symmetries, to be either gapless or gapped with topological orders, no matter how strong the interactions are~\cite{wang2014interacting}. Therefore, the impossibility for a symmetric gapped 2D surface state without topological order leads to the the topological obstruction for gapped SC and charge orders of strongly interacting electrons in 2D flat bands, which supersedes the Chern-vorticity theorem at the mean-field level.

To understand the above claims, let us first establish the map between 2D flat bands and surface states of 3D SPTs. In the non-interacting limit, a 3D class AIII SPT with bulk invariant $\nu\in\mathbb{Z}$ can host $|\nu|$ massless Dirac fermions on its surface. Focusing on $\nu=\pm 1$ cases, the massless surface modes can be depicted by Dirac Hamiltonian
\begin{equation}
\label{eq:surfdirac}
h_{\nu=\pm 1}=\sum_{\mathbf{k}}\psi_{\text{surf},\mathbf{k}}^{\dag}\hbar v(k_{x}\sigma_{x}\mp k_{y}\sigma_{y})\psi_{\text{surf},\mathbf{k}}
\end{equation}
where $\psi^{\dag}_{\text{surf},\mathbf{k}}=(\psi^{\dag}_{\text{surf},1,\mathbf{k}},\psi^{\dag}_{\text{surf},2,\mathbf{k}})$ and  $\psi^{\dag}_{\text{surf},l,\mathbf{k}}$ creates an electron with momentum $\mathbf{k}$ and orbital index $l=1,2$ on the surface; and $\sigma_{x,y,z}$ are Pauli matrices. $h_{\nu=\pm 1}$ preserves an anti-unitary particle-hole (chiral) symmetry $\hat{S}^{\prime}$: 
\begin{equation}
\hat{S}^{\prime}\psi^{\dag}_{\text{surf},l,\mathbf{k}}\hat{S}^{\prime -1}=(\sigma_{z})_{ll^{\prime}}\psi_{\text{surf},l^{\prime},\mathbf{k}}, \ \hat{S}^{\prime}i \hat{S}^{\prime -1}=-i.
\end{equation}
Upon applying a magnetic field $B$ as illustrated in Fig.~\ref{fig:illustration}(b), the surface Dirac mode develops Landau levels (LLs), the zeroth of which is pinned at zero energy due to the chiral symmetry. Several properties of the zeroth LL that are important for our following analysis, as we list below (see SM\cite{supp} for details): (i) The zeroth LL wave function is identical to a 2D lowest LL (LLL). (ii) The states on the zeroth LL, denoted as $d^{\dag}_{\text{ZLL}, n}$ where $n$ labels the orbitals, transform under the chiral symmetry as $\hat{S}^{\prime}d^{\dag}_{\text{ZLL},n}\hat{S}^{\prime -1}=\sgn(\nu)\sgn(B)d_{\text{ZLL},n}$. (iii) the Chern number of the zeroth LL of is given by $\sgn(B)$. Given point (i), mapping a flat Chern band to a surface zeroth LL is equivalent to mapping a flat Chern band to a 2D LLL. The map from a Chern band of Chern number $\mathcal{C}$ to a 2D LLL can be implemented by mapping each Bloch-Wannier state~\cite{qi2011generic}, which is localized along one direction but extended along the other direction, to the stripe-like LLL eigenstate in the Landau gauge. Specifically, the creation operator of a Bloch-Wannier state takes the form of 
\begin{equation}
\label{eq:Bloch-Wannier}
c_{k_{y},x_{0}}^{\dag}=\int d^2 r W_{k_{y},x_{0}}(\mathbf{r})\psi_{\mathbf{r}}^{\dag},
\end{equation}
where $\psi^{\dag}_{\mathbf{r}}$ creates an electron at position $\mathbf{r}$, and $W_{k_{y},x_{0}}(\mathbf{r})$ is the Bloch-Wannier wave function centered at $x_{0}$ along $x$-direction. $W_{k_{y},x_{0}}(\mathbf{r})$ can be generally derived by Fourier transforming the Bloch wave function $\Phi_{\mathbf{k}}(\mathbf{r})$ along $x$-direction: $W_{k_{y},x_{0}}(\mathbf{r})=\sum_{k_{x}}e^{-i k_{x} x_0}\Phi_{\mathbf{k}}(\mathbf{r})/\sqrt{N_{x}}$, where $N_{x}$ is the number of unit cells along $x$-direction.  Replacing $W_{k_{y},x_{0}}(\mathbf{r})$ by the Gaussian function $\frac{1}{\sqrt{\pi^{1 / 2} L_{y} l_B}}e^{i k_{y} y} e^{-\left(x- \mathcal{C} k_{y} l_B^2 \right)^2 / 2 l_B^2}$ maps the Bloch-Wannier state to a stripe-like state in LLL. Here, $L_{y}$ is the linear size of the system along $y$-direction, $l_{B}=\sqrt{\hbar/eB}$ is the magnetic length, and $k_{y}=2\pi n/L_{y}$ with $n$ being an integer.  Note that for the stripe-like LLL state, $x_{0}$ and $k_{y}$ are both determined by the integer $n$ in the range of $[1,eBL_{x}L_{y}/h]$, where $L_{x}$ is the linear size of the system along $x$-direction and $eBL_{x}L_{y}/h$ is the Landau degeneracy. Hereafter, we denote the creation/annihilation operator of stripe-like LLL state by $c^{\dag}_{n}/c_{n}$. 

After establishing the map, we are now ready to discuss the topological obstructions for the SC and charge orders. As depicted in Fig.~\ref{fig:illustration}(b), we map the flat Chern bands with Chern number $C_{\pm}$ to the surface states of 3D SPTs with bulk topological invariants $\nu_{\pm}$ and subjected to magnetic field $B_{\pm}$. Meanwhile, we map the 2D chiral symmetry $\hat{S}$ to the surface symmetry $\hat{S}^{\prime}$, which requires $\sgn(\nu_{+})\sgn(B_{+})=-\sgn(\nu_{-})\sgn(B_{-})=+1$ according to aforementioned property (ii). Note that although $\hat{S}^\prime$ originates from a locality-preserving (on-site) symmetry of the 3D system, due to the nontrivial bulk topology of the 3D SPT, it is mapped to a nonlocal symmetry $\hat{S}$ in 2D. For the SC order where $U_{c}(1)$ is spontaneously broken, $U_{v}(1)\times Z_2^{\hat{S}^{\prime}}$ symmetry leads to a 3D bulk characterized by a topological invariant $\nu_{v}=\nu_{+}-\nu_{-}$. Consequently, when $\nu_{+}=-\nu_{-}=+1$ and thus $B_{+}=B_{-}$, we have $\nu_{v}=2$ indicating that the surface cannot be symmetrically gapped out unless developing a topological order. This directly implies that when the two flat Chern bands have $\mathcal{C}_{+}=\mathcal{C}_{-}=+1$ [c.f. point (iii)], there cannot be symmetric, gapped SC phase without topological order, i.e., the SC order is topologically obstructed. For the charge order where $U_{v}(1)$ is spontaneously broken, the 3D bulk will be characterized by a topological invariant $\nu_{c}=\nu_{+}+\nu_{-}$ with $U_{c}(1)\times Z_2^{\hat{S}^{\prime}}$ symmetry. Similarly, when $\nu_{+}=\nu_{-}=+1$ and thus $B_{+}=-B_{-}$, the $\nu_{c}$ is 2 which indicates that when $\mathcal{C}_{+}=-\mathcal{C}_{-}=1$, the charge order is topologically obstructed. 


The above arguments for topological obstructions is valid independent of the form and the strength of interactions we consider, as long as the $U(1)\times Z_2^{\hat{S}^{\prime}}$ symmetry is preserved. Naively, the interacting topological obstruction should be consistent with the Chern-vorticity theorem derived from the band theory, corresponding to a $\mathbb{Z}$ classification associated with the valley/charge Chern number $\mathcal{C}_{c/v}\in\mathbb{Z}$. However, as we will discuss below, strong interactions give rise to a $\mathbb{Z}_8$ topological obstruction, in contrast to the $\mathbb{Z}$ obstruction at the non-interacting level. 

\magenta{\it $\mathbb{Z}_{8}$ classification of the topological obstruction.|}
In the strongly interacting regime, it is known that the classification of the 3D fermion SPT of class AIII reduces from $\mathbb{Z}$ to $\mathbb{Z}_{8}$\cite{wang2014interacting}. This means that the surface states can be gapped out symmetrically without developing any topological orders when $\nu=8$. Consequently, if the net Chern numbers in $\pm$ valleys are $\mathcal{C}_{+}=\mathcal{C}_{-}=4$ ($\mathcal{C}_{+}=-\mathcal{C}_{-}=4$)\footnote{This is equivalent to stacking four copies of those depicted in Fig.~\ref{fig:illustration}.}, the SC (charge) order is no longer topologically obstructed. Since the chiral symmetry is nonlocal in 2D flat Chern bands, it is difficult to explicitly write down a chiral symmetric ground state. The simplest way to understand the $\mathbb{Z}_{8}$ classification is to the vortex proliferation argument for 3D TI surface discussed in Ref.~\cite{wang2014interacting}. For pedagogical purposes, we briefly review it here and then apply it to the SC and charge orders. 

Let us first consider the charge order where the 3D bulk invariants associated with $U_c(1)\times Z_2^{\hat{S}^\prime}$ symmetry are $\nu_{+}=\nu_{-}=+4$ and thus $\nu_{c}=8$, corresponding to flat Chern bands with $\mathcal{C}_{+}=-\mathcal{C}_{-}=+4$. In this case, there are eight Dirac modes on the surface, each of which is described by $h_{\nu=+1}$ in Eq.~\eqref{eq:surfdirac}. If breaking the $U_{c}(1)$ symmetry protecting the bulk topology, the surface states can be gapped out by the following pairing term
\begin{equation}
\label{eq:surfacepairing}
h_{1}=\sum_{j,\mathbf{k}}\left(\psi_{\text{surf},j,\mathbf{k}}\right)^{T}i\Delta\sigma_{y}\psi_{\text{surf},j,\mathbf{k}}+h.c.,
\end{equation}
where $j=(\pm,m)$ with valley index $\pm$ and flavor index $m=1,2,3,4$, and $\Delta$ is the pairing order parameter. Though breaking the $U_{c}(1)$ symmetry, this term preserves the chiral symmetry $\hat{S}^{\prime}$. Strong interactions can lead to quantum fluctuations of $\Delta$, and restore $U_{c}(1)$ symmetry by proliferating vortices of the SC order parameter $\Delta$. To understand the vortex porliferation process, we need to examine the zero-energy modes in each vortex. Each surface Dirac mode produces a Majorana zero mode (MZM), $\gamma_{n}$, in one vortex~\cite{fu2008superconducting} that transforms trivially under the chiral symmetry: $\hat{S}^{\prime}\gamma_{n}\hat{S}^{\prime -1}=\gamma_{n}$. Consequently, eight surface Dirac modes leads to eight MZMs in each vortex, which can be gapped out symmetrically~\cite{fidkowski2010effects}. Given that the symmetrically gapped vortex state is a linear combination of states with even fermion parity, each vortex is a bosonic excitation~\cite{wang2014interacting,supp} and consequently we can condense the vortices to restore $U_{c}(1)$ symmetry, leaving the surface to be a trivial, symmetric, gapped state. The same argument applies to four copies of topologically obstructed SC orders with $\nu_{v}=8$, and we discuss the details in SM~\cite{supp}.


Following a similar argument, one can show that the surface can be symmetrically gapped out and develop a $\mathbb{Z}_2$ topological order when $\nu=4$~\cite{wang2014interacting}. This suggests that a symmetric, gapped SC (charge) order in flat Chern bands with $\mathcal{C}_{+}=\mathcal{C}_{-}=+2$ ($\mathcal{C}_{+}=-\mathcal{C}_{-}=+2$) must coexist with a topological order. Below we construct such an interacting Hamiltonian and identify its symmetric ground state with coexisting SC and topological orders.

\magenta{\it Symmetrically gapped SC state with coexisting topological order.|} In this section we construct a gapped SC state coexisting with topological order, formulated using Wannier-Bloch states of the flat Chern bands. To see this, we consider two flat bands labelled by index $j=1,2$ in each of the $\pm$ valleys, each with a Chern number of $+1$, corresponding to $\mathcal{C}_{+}=\mathcal{C}_{-}=+2$. As discussed previously, a mean-field pairing term $H_{\text{pair}}=\sum_{\mathbf{k},j=1,2}(\Delta_{\rm SC}(\mathbf{k})c_{+,j,\mathbf{k}}^{\dag}c_{-,j,-\mathbf{k}}^{\dag}+h.c.)$ cannot gap out the systems because $\Delta_{\rm SC}(\mathbf{k})$ must vanish at certain momenta. Other interactions are needed to achieve a gapped superconductor, and we consider the hard-core interaction
\begin{equation}
\label{eq:two-body0}
H_{\text{int}}=\int d^{2}r d^{2}r^{\prime} V(\mathbf{r}-\mathbf{r}^{\prime})\psi^{ \dag}_{+,1,\mathbf{r}}\psi^{ \dag}_{-,1,\mathbf{r}^{\prime}}\psi_{+,2,\mathbf{r}^{\prime}}\psi_{-,2,\mathbf{r}} +h.c.
\end{equation}
where $V(\mathbf{r}-\mathbf{r}^{\prime})=V_{0}\delta(\mathbf{r}-\mathbf{r}^{\prime})$ and $\psi^{\dag}_{\pm,m,\mathbf{k}}$ creates fermions in the $m$-th flavor of valley $\pm$. $H_{\text{int}}$ preserves the full symmetry $U_{c}(1)\times U_{v}(1)\times \hat{S}$, though $U_c(1)$ is explicitly broken by $H_\text{pair}$. Projecting the Hilbert space onto the flat Chern bands, the fermion creation operator can be expanded by a linear combination of Bloch-Wannier states:
\begin{equation}
\label{eq:LL}
\psi^{\dag}_{\pm,m,\mathbf{r}}\approx \sum_{(k_{y},x_{0})}W^{\star}_{\pm,m,k_{y},x_{0}}(\mathbf{r})c^{\dag}_{k_{y},x_{0}}.
\end{equation}
Without loss of generality, we use the LLL form $W_{\pm,m,k_{y},x_{0}}(\mathbf{r})=e^{i k_{y}y}\frac{1}{\sqrt{\pi^{1 / 2} L_{y} l_B}} e^{-\left(x-  \mathcal{C}_{\pm}k_{y} l_B^2 \right)^2 / 2 l_B^2}$, for which the label $(k_{y},x_{0})$ can be combined into an integer index $n$ as discussed previously. Substituting Eq.~\eqref{eq:LL} into Eq.~\eqref{eq:two-body0}, we have 

\begin{equation}
\label{eq:hint}
H_{\text{int}}=\sum_{l,p,q}\tilde{V}_{p,q}
c^{\dag}_{+,1,l+p}c^{\dag}_{-,1,l-p}c_{+,2,l-q}c_{-,2,l+q}+h.c.,
\end{equation}
where $\tilde{V}_{p,q}=g\kappa\exp\left\{-\kappa^2(p^2+q^2)\right\}$ with $\kappa=2\pi l_{B}/L_{y}$ and $g=V_{0}/[(2\pi)^{3/2}l_{B}^2]$. $l,p,q$ take half integer values, and $l\pm q$ and $l\pm p$ are integers labelling the center of  the stripe-like states along $x$-direction. From Eq.~\eqref{eq:two-body0} to Eq.~\eqref{eq:hint}, we reduce a 2D problem into a 1D problem by rewriting the two-body potential in terms of stripe-like states in LLL~\cite{lee2004mott,seidel2005incompressible}. Details can be found in SM~\cite{supp}.

Let us first consider the thin cylinder limit, where $x$-direction is open, and $y$-direction is periodic with $ L_{y}\ll l_{B}$. In this limit, $\kappa$ is large, and Eq.~\eqref{eq:hint} is dominated by its leading term 
\begin{equation}
\label{eq:hint1}
\tilde{H}_{\text{int}}=g\kappa\sum_{l} c^{\dag}_{+,1,l}c^{\dag}_{-,1,l}c_{+,2,l}c_{-,2,l}+h.c.,
\end{equation}
which has a ground state at half-filling with energy $-g\kappa$:
\begin{equation}
\label{eq:G0}
\ket{G_{0}}=\prod_{l}(c^{\dag}_{+,1,l}c^{\dag}_{-,1,l}-c^{\dag}_{+,2,l}c^{\dag}_{-,2,l})\ket{0}.
\end{equation}
Since $\hat{S}\ket{0}=c^{\dag}_{+,1,l}c^{\dag}_{-,1,l}c^{\dag}_{+,2,l}c^{\dag}_{-,2,l}\ket{0}$, it is straightforward to verify that $\ket{G_{0}}$ preserves $U_{c}(1)\times U_{v}(1)\times \hat{S}$ symmetry. Assuming weak pairing, we can treat $H_{\text{pair}}$ as a perturbation to $\tilde{H}_{\text{int}}$, and a gapped SC state can be obtained using non-degenerate perturbation theory:
\begin{equation}
\label{eq:G}
\ket{G}=\ket{G_{0}}-\sum_{\text{exc}}\frac{\bra{\text{exc}}H_{\text{pair}}\ket{G_{0}}}{E_{\text{exc}}+g\kappa}\ket{\text{exc}}+\cdots
\end{equation}
In the thin cylinder limit, all excited states $\ket{\text{exc}}$ with non-vanishing coefficient in Eq.~\eqref{eq:G} must have $E_{\text{exc}}=0$. 
Importantly, $\ket{G}$ is $U_{v}(1)\times Z_2^{\hat{S}}$ symmetric because $H_{\text{pair}}$, $\ket{G_{0}}$, and $\sum_{\text{exc}}\ket{\text{exc}}\bra{\text{exc}}=\mathbbm{1}-\ket{G_{0}}\bra{G_{0}}$ all preserve $U_{v}(1)\times \hat{S}$ symmetric~\cite{supp}. Since $H_{\text{pair}}$ is only a perturbation and the energy gap of excitations is dominated by $g\kappa$, the fact that $\Delta(\mathbf{k})$ vanishes at some momenta will not affect the gapped nature of the ground state. 

\begin{figure}[h]
    \centering
    \includegraphics[width=0.48\textwidth]{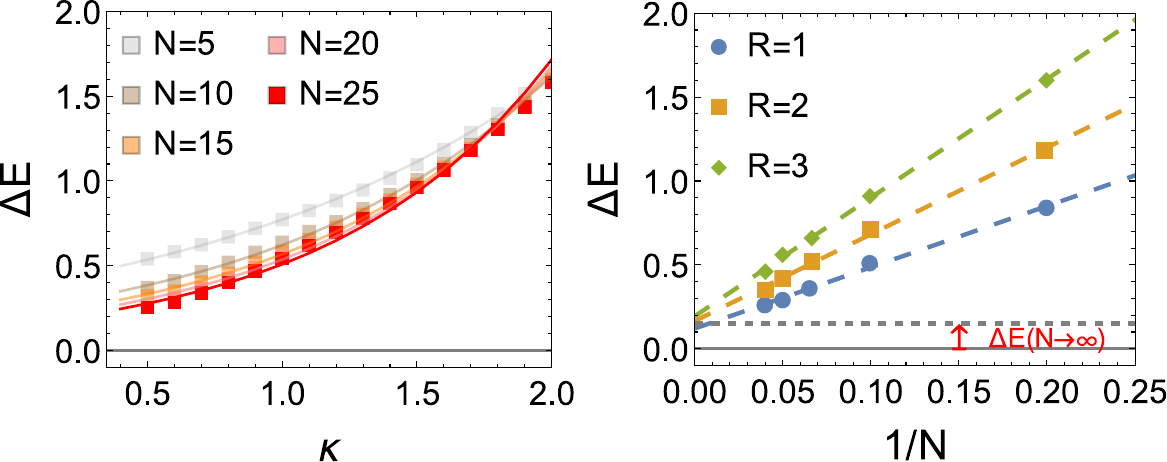}
    \caption{Energy Gap $\Delta E$ (in unit of $g$) of Eq.~\eqref{eq:hint1} for different sizes ($N$) and geometries. (left) Gaps under different $\kappa$ and different $N$, and their fittings $\Delta E(\kappa; N)$ by exponential functions. (right) Finite size scaling of gaps as a function of $1/N$, under fixed aspect rations ($R = 1,2,3$). Data points in markers are obtained by DMRG under open boundary condition, with the maximal bond dimension $d=800$ and truncation errors $\epsilon \sim 10^{-7}$. Details can be found in SM \cite{supp}. }
    \label{fig:dmrg}
\end{figure}

Based on the $\mathbb{Z}_8$ classification discussed previously, a gapped, symmetric $\ket{G}$ in the thin cylinder limit indicates a ground state where SC order and topological order coexist, provided the gap does not close as we increase $L_{y}$ (decrease $\kappa$) and the system size approaches the 2D limit. Next, we provide numerical evidence that the gap will indeed persist in a 2D cylinder geometry. 
Specifically, we perform density matrix renormalization group (DMRG) calculations \cite{itensor,supp} for the effective 1D model \eqref{eq:hint} for various $\kappa=2\pi l_b/L_y$ and system sizes $N=L_xL_y$. As $\kappa$ decreases for systems of different sizes, no level crossing has been observed. As shown in Fig.~\ref{fig:dmrg}(left), with fixed system size $N$, the gap will saturate to a finite value as $\kappa$ decreases. This saturation value of the energy gap converges as we increase the system size. Moreover, we present the finite-size scaling of the energy gap for systems with fixed aspect ratios $R \equiv L_x/L_y = 1,2,3$ in Fig.~\ref{fig:dmrg}(right). This amounts to extrapolating to the true 2D limit with $\kappa \rightarrow 0, ~L_x,L_y \sim O(\kappa^{-1})=O(\sqrt{N})$. The energy gap of systems with different $R$ converges to the same finite value in the thermodynamic limit, i.e., $\Delta E(1/N \rightarrow 0) \sim 0.15 g$, as marked out by the gray dashed horizontal line in Fig.~\ref{fig:dmrg}(right). These numerical evidences strongly suggest that no phase transition occurs as the system evolves from a thin cylinder to a 2D cylinder geometry. Consequently, as we increase $L_y$, $\ket{G}$ will remain a gapped, symmetric ground state with SC order in 2D. As argued previously, such a gapped symmetric state with $\nu_v=4$ must also exhibit topological orders, in addition to SC order. Therefore we reach a symmetric gapped ground state with coexisting long-range and topological orders. 


\magenta{\it Discussions|}
In this work, we have established an LSM-type topological obstruction to trivially gapped SC and charge orders in strongly correlated flat Chern bands at half filling. Our results reveal an intriguing connection between long-range orders and topological orders, i.e., the presence of certain symmetric, gapped long-range orders is necessarily accompanied by topological orders. Since the Landau-type long-range orders are much easier to detect in experiments than topological orders, the topological obstructions discussed in this work may serve as a indicator for the presence of topological orders. Importantly, the flat Chern bands discussed in this work can be realized either in 2D Morie systems~\cite{zhang2019nearly,chen2020tunable,bultinck2020mechanism,serlin2020intrinsic} and/or on the Fermi surface of 3D topological semimetals such as Weyl semimetals~\cite{Li2018topological,armitage2018weyl,lv2021experimental}, for which the $U_{v}(1)$ symmetry are indeed realized as a valley-$U(1)$ symmetry. We emphasize that the emergent chiral symmetry, arising from projecting the Hilbert space onto flat bands, is vital for the topological obstruction. If this symmetry is broken, a gapped incommensurate valley coherence (IVC) order can emerge. In twisted bilayer-graphenes, the finite width and nontrivial dispersion of the flat Chern bands will generally break the emergent chiral symmetry. As a result, incommensurate Kekul\'{e} orders have been theoretically proposed~\cite{kwan2021kekule,wang2024chern,kwan2024textured} experimentally confirmed~\cite{nuckolls2023quantum}. The emergent chiral symmetry of flat bands in real materials should be considered as an approximate symmetry. The flatter the bandwidth relative to the interaction strength, the more accurate this approximation becomes. Therefore, we expect the topological obstruction discussed in our paper would be more influential in systems with flatter Chern bands, e.g. in twisted MoTe$_2$ homobilayers~\cite{wu2019topological,devakul2021magic,dong2023composite,crepel2024chiral}, where the breaking of the emergent chiral symmetry is expected to be weaker and thus it is harder to have a trivially gapped IVC order. 


Looking forward, two questions naturally arise from this work: (i) can one explicitly construct a featureless and gapped ground state (w/o topological orders) for the $\nu=8$ case, e.g. in the context of SC orders in topological flat bands with $\mathcal{C}_+=\mathcal{C}_-=4$? (ii) Can the topological obstructions discussed here be generalized to correlated flat bands of other symmetry classes, such as class DIII and class CII? We leave these intriguing questions for future works. 
\\

\begin{acknowledgments}
\textbf{Acknowledgments: }
We thank Jiabin Yu, Ya-Hui Zhang, and Rui-Xing Zhang for their insightful discussions. We also extend our gratitude to Siddharth A. Parameswaran and Nemin Wei for their valuable discussions on materials. This research was primarily supported by the Center for Emergent Materials, an NSF MRSEC, under award number DMR-2011876.
\end{acknowledgments}

\bibliography{refs.bib}




\end{document}


\title{Supplemental Material for ``\textbf{\ourtitle}"}

\author{Penghao Zhu} 
\author{Shi Feng}
\author{Yuan-Ming Lu}
\affiliation{Department of Physics, The Ohio State University, Columbus, OH 43210, USA}


\maketitle
\tableofcontents

\setcounter{section}{0}
\setcounter{figure}{0}
\setcounter{equation}{0}
\renewcommand{\thefigure}{S\arabic{figure}}
\renewcommand{\theequation}{S\arabic{equation}}
\renewcommand{\thesection}{S\arabic{section}}
\onecolumngrid



\section{Chiral symmetry of two-body interactions}
We demonstrate that the chiral symmetry $\hat{S}$, defined by
\begin{equation}
\label{eq
}
\hat{S}c^{\dag}_{\pm,m,\mathbf{k}}\hat{S}^{-1}=\pm c_{\pm,m,\mathbf{k}}, \quad \hat{S} i \hat{S}^{-1}=-i,
\end{equation}
is generally preserved up to a one-body term for two-body interactions involving \textit{an even number} of fermionic operators in both the $+$ and $-$ valleys. Note that $\pm$ label the valleys, and $m$ labels different flavors in each valley. For brevity, we use subscript $j=(\pm,m)$ in the following discussion. A general two-body interaction can then be expressed as
\begin{equation}
\label{eq:twobody}
H_{\text{2-body}}=\frac{1}{2}\sum_{j_{1}\mathbf{k}_1,j_{2}\mathbf{k}_2} \sum_{j_{1}^{\prime}\mathbf{k}_1^{\prime},j_{2}^{\prime}\mathbf{k}_2^{\prime}}V_{j_{1}^{\prime}\mathbf{k}_1^{\prime},j_{2}^{\prime}\mathbf{k}_2^{\prime};j_{1}\mathbf{k}_1,j_{2}\mathbf{k}_2}c_{j_{1}^{\prime}\mathbf{k}_1^{\prime}}^{\dag}c_{j_{2}^{\prime}\mathbf{k}_2^{\prime}}^{\dag}c_{j_{2}\mathbf{k}_2}c_{j_{1}\mathbf{k}_1}+h.c..
\end{equation}
Given that the number of fermionic operators in the $+$ ($-$) valley is even, under the chiral symmetry, the first term of Eq.~\eqref{eq:twobody} becomes
\begin{equation}
\label{eq:chiraltrans}
\hat{S}V_{j_{1}^{\prime}\mathbf{k}_1^{\prime},j_{2}^{\prime}\mathbf{k}_2^{\prime};j_{1}\mathbf{k}_1,j_{2}\mathbf{k}_2}c_{j_{1}^{\prime}\mathbf{k}_1^{\prime}}^{\dag}c_{j_{2}^{\prime}\mathbf{k}_2^{\prime}}^{\dag}c_{j_{2}\mathbf{k}_2}c_{j_{1}\mathbf{k}_1}\hat{S}^{-1}=V^{\star}_{j_{1}^{\prime}\mathbf{k}_1^{\prime},j_{2}^{\prime}\mathbf{k}_2^{\prime};j_{1}\mathbf{k}_1,j_{2}\mathbf{k}_2}c_{j_{1}^{\prime}\mathbf{k}_1^{\prime}}c_{j_{2}^{\prime}\mathbf{k}_2^{\prime}}c^{\dag}_{j_{2}\mathbf{k}_2}c^{\dag}_{j_{1}\mathbf{k}_1},
\end{equation}
where 
\begin{equation}
\begin{aligned}
c_{j_{1}^{\prime}\mathbf{k}_1^{\prime}}c_{j_{2}^{\prime}\mathbf{k}_2^{\prime}}c^{\dag}_{j_{2}\mathbf{k}_2}c^{\dag}_{j_{1}\mathbf{k}_1}&=-\delta_{j_{2}^{\prime}\mathbf{k}_2^{\prime},j_{1}\mathbf{k}_1}c_{j_{1}^{\prime}\mathbf{k}_1^{\prime}}c^{\dag}_{j_{2}\mathbf{k}_2}+\delta_{j_{1}^{\prime}\mathbf{k}_1^{\prime},j_{1}\mathbf{k}_1}c_{j_{2}^{\prime}\mathbf{k}_2^{\prime}}c^{\dag}_{j_{2}\mathbf{k}_2}-\delta_{j_{2}^{\prime}\mathbf{k}_2^{\prime},j_{2}\mathbf{k}_2}c^{\dag}_{j_{1}\mathbf{k}_1}c_{j_{1}^{\prime}\mathbf{k}_1^{\prime}}
\\
&+\delta_{j_{1}^{\prime}\mathbf{k}_1^{\prime},j_{2}\mathbf{k}_2}c^{\dag}_{j_{1}\mathbf{k}_1}c_{j_{2}^{\prime}\mathbf{k}_2^{\prime}}+c^{\dag}_{j_{1}\mathbf{k}_1}c^{\dag}_{j_{2}\mathbf{k}_2}c_{j_{2}^{\prime}\mathbf{k}_2^{\prime}}c_{j_{1}^{\prime}\mathbf{k}_1^{\prime}}.
\end{aligned}
\end{equation}
Since $V^{\star}_{j_{1}^{\prime}\mathbf{k}_1^{\prime},j_{2}^{\prime}\mathbf{k}_2^{\prime};j_{1}\mathbf{k}_1,j_{2}\mathbf{k}_2}=V_{j_{1}\mathbf{k}_1,j_{2}\mathbf{k}_2;j_{1}^{\prime}\mathbf{k}_1^{\prime},j_{2}^{\prime}\mathbf{k}_2^{\prime}}$ and the second term is the Hermitian conjugation of the first term, the 2-body interactions transform as
\begin{equation}
\label{eq:twobodytrans}
\hat{S}H_{\text{2-body}}\hat{S}^{-1}=H_{\text{2-body}}+H_\text{{1-body}},
\end{equation}
where 
\begin{equation}
\label{eq:onebody}
\begin{aligned}
H_{\text{1-body}}&=\frac{1}{2}\sum_{j_{1}\mathbf{k}_1,j_{2}\mathbf{k}_2} \sum_{j_{1}^{\prime}\mathbf{k}_1^{\prime},j_{2}^{\prime}\mathbf{k}_2^{\prime}}V^{\star}_{j_{1}^{\prime}\mathbf{k}_1^{\prime},j_{2}^{\prime}\mathbf{k}_2^{\prime};j_{1}\mathbf{k}_1,j_{2}\mathbf{k}_2}(-\delta_{j_{2}^{\prime}\mathbf{k}_2^{\prime},j_{1}\mathbf{k}_1}c_{j_{1}^{\prime}\mathbf{k}_1^{\prime}}c^{\dag}_{j_{2}\mathbf{k}_2}+\delta_{j_{1}^{\prime}\mathbf{k}_1^{\prime},j_{1}\mathbf{k}_1}c_{j_{2}^{\prime}\mathbf{k}_2^{\prime}}c^{\dag}_{j_{2}\mathbf{k}_2}
\\
& \ \ \ \ \ \ \ \ \ \ \ \ \ \ \ \ \ \ \ \ \ \ \ \ \ \ \ \ \ \ \ -\delta_{j_{2}^{\prime}\mathbf{k}_2^{\prime},j_{2}\mathbf{k}_2}c^{\dag}_{j_{1}\mathbf{k}_1}c_{j_{1}^{\prime}\mathbf{k}_1^{\prime}}+\delta_{j_{1}^{\prime}\mathbf{k}_1^{\prime},j_{2}\mathbf{k}_2}c^{\dag}_{j_{1}\mathbf{k}_1}c_{j_{2}^{\prime}\mathbf{k}_2^{\prime}})+h.c.
\\
&=\frac{1}{2}\sum_{j_{1}\mathbf{k}_1,j_{2}\mathbf{k}_2} (V^{\star}_{j_{1}\mathbf{k}_1,j_{2}\mathbf{k}_2;j_{1}\mathbf{k}_1,j_{2}\mathbf{k}_2}-V^{\star}_{j_{2}\mathbf{k}_2,j_{1}\mathbf{k}_1;j_{1}\mathbf{k}_1,j_{2}\mathbf{k}_2}) +\frac{1}{2}\sum_{j_{1}\mathbf{k}_1,j_{2}\mathbf{k}_2,j^{\prime}\mathbf{k}^{\prime}}(V^{\star}_{j_{1}\mathbf{k}_{1},j^{\prime}\mathbf{k}^{\prime};j^{\prime}\mathbf{k}^{\prime},j_{2}\mathbf{k}_{2}}
\\
& \ \ \ \ \ \ \ \ -V^{\star}_{j^{\prime}\mathbf{k}^{\prime},j_1\mathbf{k}_1;j^{\prime}\mathbf{k}^{\prime},j_{2}\mathbf{k}_{2}}-V^{\star}_{j_1\mathbf{k}_1,j^{\prime}\mathbf{k}^{\prime};j_{2}\mathbf{k}_{2},j^{\prime}\mathbf{k}^{\prime}}+V^{\star}_{j^{\prime}\mathbf{k}^{\prime},j_1\mathbf{k}_1;j_{2}\mathbf{k}_{2},j^{\prime}\mathbf{k}^{\prime}})c_{j_{2}\mathbf{k}_{2}}^{\dag}c_{j_{1}\mathbf{k}_{1}}+h.c.
\end{aligned}
\end{equation}
Note that $j_{1}$ and $j_{2}$ in the last line of the above equation must be in the same valley because we have required that the number of fermionic operators in each valley is even. Then, it is straightforward to verify that $\hat{S}H_{1-body}\hat{S}^{-1}=-H_{1-body}$. Consequently, we can generally have a chiral symmetric interaction term that takes the form:
\begin{equation}
\tilde{H}=H_{\text{2-body}}+\frac{1}{2}H_{\text{1-body}}.
\end{equation}
In realistic cases, the interaction can be a Coulomb interaction such that 
\begin{equation}
\label{eq:coulomb}
   V_{j_{1}^{\prime}\mathbf{k}_{1}^{\prime},j_{2}^{\prime}\mathbf{k}_{2}^{\prime};j_{1}\mathbf{k}_{1},j_{2}\mathbf{k}_2}=\int d\mathbf{r}_{1}d\mathbf{r}_{2}\Phi^{\star}_{j_{1}^{\prime}\mathbf{k}_{1}^{\prime}}(\mathbf{r}_{1})\Phi^{\star}_{j_{2}^{\prime}\mathbf{k}_{2}^{\prime}}(\mathbf{r}_{2})\frac{e^2}{|\mathbf{r}_{1}-\mathbf{r}_{2}|}\Phi_{j_{1}\mathbf{k}_1}(\mathbf{r}_{1})\Phi_{j_{2}\mathbf{k}_2}(\mathbf{r}_{2}), 
\end{equation}
where $\Phi_{j\mathbf{k}}(\mathbf{r})$ represents the Bloch state. It is then straightforward to see $V_{j_{1}^{\prime}\mathbf{k}_{1}^{\prime},j_{2}^{\prime}\mathbf{k}_{2}^{\prime};j_{1}\mathbf{k}_{1},j_{2}\mathbf{k}_2}=V_{j_{2}^{\prime}\mathbf{k}_{2}^{\prime},j_{1}^{\prime}\mathbf{k}_{1}^{\prime};j_{2}\mathbf{k}_{2},j_{1}\mathbf{k}_1}$. Moreover, the momentum is conserved so $\mathbf{k}_{1}^{\prime}+\mathbf{k}_{2}^{\prime}=\mathbf{k}_{1}+\mathbf{k}_2$. As a result, the quadratic term in Eq.~\eqref{eq:onebody} can be simplified into
\begin{equation}
\label{eq:quadratic}
\sum_{j_{1}\mathbf{k},j_{2}\mathbf{k},j^{\prime}\mathbf{k}^{\prime}}(V^{\star}_{j_{1}\mathbf{k},j^{\prime}\mathbf{k}^{\prime};j^{\prime}\mathbf{k}^{\prime},j_{2}\mathbf{k}}-V^{\star}_{j^{\prime}\mathbf{k}^{\prime},j_1\mathbf{k};j^{\prime}\mathbf{k}^{\prime},j_{2}\mathbf{k}})c_{j_{2}\mathbf{k}}^{\dag}c_{j_{1}\mathbf{k}}+h.c.
\end{equation}
Specifically, if we consider $j_{1}=j_{2}=j^{\prime}$, i.e., interactions within one flat Chern band, this one-body term yields the Hartree-Fock single-particle energy for each state labeled by $\mathbf{k}$. For Coulomb interactions between stripe-like states in the LLL, i.e., replacing the subscript $\mathbf{k}$ by $n$ that labels the center of the stripe-like state, the Hatree-Fock single particle energy should be independent of $n$ because of the translation symmetry along the direction where the stripe like states are localized. Therefore, the single particle term is nothing but a trivial chemical potential term in this case~\cite{wang2016half}.

\section{Review of Chern-vorticity theorem}

\begin{figure}[h]
    \centering
    \includegraphics[width=0.8\columnwidth]{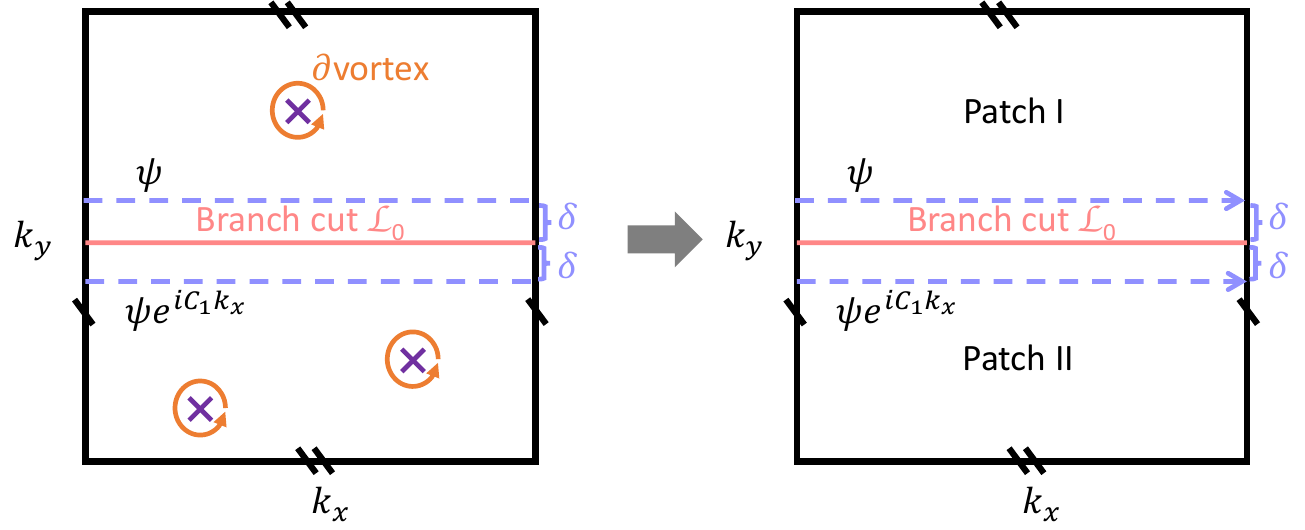}
    \caption{Illustration of the vortices, two gauge patches over a 2D BZ, and the paths for the line integrals. The purple crosses represent vortices of $\mathcal{O}(\mathbf{k}_{1}
    )$. The upper (lower) blue dashed line represents $\mathcal{L}_{0}+\delta$ ($\mathcal{L}_{0}-\delta$).}
    \label{fig:chernvorticity}
\end{figure}

Here, we study the vortices of SC and charge order pairings projected onto two bands in opposite valleys (as discussed in the main text):
\begin{equation}
    {\cal O}({\bf k}_1) = \bra{\psi_{1}(\mathbf{k}_1)}\hat{O}\ket{\psi_{2}(\mathbf{k}_{2})} = |{\cal O}({\bf k}_1)|e^{i \varphi({\bf k}_1)},
\end{equation}
where $\hat{O}$ can be the SC pairing term or the charge order pairing term. We focus on pairs with zero net momentum. Consequently, for SC pairing we consider $\mathbf{k}_2=-\mathbf{k}_{1}$, while for charge order we consider $\mathbf{k}_2=\mathbf{k}_{1}$.  For clarity, we denote the Bloch states of the electron bands 1 and 2 to be $\ket{\xi_{1}^{(e)}}$ and $\ket{\xi_{2}^{(e)}}$, respectively. If $\hat{O}$ is the SC pairing (charge order pairing), we can choose $\ket{\psi_{1}(\mathbf{k}_{1})}=\ket{\xi^{(e)}_{1}(\mathbf{k}_{1})}$ and $\ket{\psi_{2}(\mathbf{k}_{2})}=\ket{\xi^{(e)\star}_{2}(\mathbf{k}_{2})}$ ($\ket{\psi_{1}(\mathbf{k}_{1})}=\ket{\xi^{(e)}_{1}(\mathbf{k}_{1})}$ and $\ket{\psi_{2}(\mathbf{k}_{2})}=\ket{\xi^{(e)}_{2}(\mathbf{k}_{2})}$) to properly project the particle-particle (particle-hole) pairing. A vortex of $\mathcal{O}(\mathbf{k}_{1})$ corresponds to a zero point of  $\mathcal{O}(\mathbf{k}_{1})$, around which $\varphi(\mathbf{k}_{1})$ has a $2\pi$-quantized winding. Specifically, the vorticity of a vortex of $\mathcal{O}(\mathbf{k}_{1})$ is defined as $(1/2\pi)\oint_{\partial\text{vortex}}\partial_{\mathbf{k}_{1}}\varphi\cdot d\mathbf{k}_{1}$. The net vorticity over a 2D manifold in momentum space can be calculated through $vort=(1/2\pi)\sum_{\text{vortex}}\oint_{\partial\text{vortex}}\partial_{\mathbf{k}_{1}}\varphi\cdot d\mathbf{k}_{1}$, as illustrated for a 2D BZ in Fig.~\ref{fig:chernvorticity}.

With these setups, we next prove the Chern-vorticity theorem: The net vorticity of $\mathcal{O}(\mathbf{k}_{1})$ is determined by the Chern numbers, $\mathcal{C}_{1}$ and $\mathcal{C}_{2}$,  of $\ket{\psi_{1}(\mathbf{k})}$ and $\ket{\psi_{2}(\mathbf{k})}$. We start by constructing a vector:
\begin{equation}
\label{eq:vector}
\mathbf{S}=\partial_{\mathbf{k}_{1}}\varphi(\mathbf{k}_{1})- \mathbf{A}_{1}(\mathbf{k}_{1})+g \mathbf{A}_{2}(\mathbf{k}_{2}),
\end{equation}
where
\begin{equation}
\label{eq:BC}
\mathbf{A}_{a}(\mathbf{k}_{a})=i\langle \psi_{a}(\mathbf{k}_{a})|\partial_{\mathbf{k}_{a}}\psi_{a}(\mathbf{k}_{a})\rangle, \ a=1,2,
\end{equation}
is the Berry connection, and $g=-1 (+1)$ for SC (charge order) pairing. Importantly, the vector $\mathbf{S}$ is invariant under the $U(1)$ gauge transformation $|\psi_{a}(\mathbf{k}_{a})\rangle\rightarrow e^{i\alpha_{a}(\mathbf{k}_{a})}|\psi_{a}(\mathbf{k}_{a})\rangle$ with $a=1,2$.
If there is a nonzero Chern-number of $\ket{\psi_{1}(\bf k_{1})}$ ($\ket{\psi_{2}(\bf k_{2})}$) on the corresponding closed 2D manifold (e.g. a 2D BZ in Fig.~\ref{fig:chernvorticity}), then $\mathbf{A}_{1}(\mathbf{k}_1)$ ($\mathbf{A}_{2}(\mathbf{k}_2)$) cannot be globally smooth over the entire manifold. Without loss of generality, one can have two locally smooth gauge patches over the 2D BZ as illustrated in Fig.~\ref{fig:chernvorticity}. Then, across the boundary (i.e., $\mathcal{L}_{0}$ in Fig.~\ref{fig:chernvorticity}) of the two patches, there are singularities of $\mathbf{A}_{1}(\mathbf{k}_1)$ and $\mathbf{A}_{2}(\mathbf{k}_2)$ such that
\begin{equation}
\label{eq:integral}
\begin{aligned}
&\lim_{\delta\rightarrow 0}\left(\int_{\mathcal{L}_{0}+\delta}-\int_{\mathcal{L}_{0}-\delta}\right)\mathbf{A}_{1}(\mathbf{k}_{1})\cdot d\mathbf{k}_{1}=2\pi \mathcal{C}_{1},
\\
& \lim_{\delta\rightarrow 0}\left(\int_{g(\mathcal{L}_{0}+\delta)}-\int_{g(\mathcal{L}_{0}-\delta)}\right)\mathbf{A}_{2}(\mathbf{k}_{2})\cdot d\mathbf{k}_{2}=2\pi\mathcal{C}_{2}.
\end{aligned}
\end{equation}

In contrast to the Berry connection, $\mathbf{S}$ is gauge invariant, and thus is continuous across the gauge patch boundary, i.e., $\lim_{\delta\rightarrow 0}\left(\int_{\mathcal{L}_{0}+\delta}-\int_{\mathcal{L}_{0}-\delta}\right)\mathbf{S}\cdot d\mathbf{k}_{1}=0$. Consequently, $\partial_{\mathbf{k}_{1}}\varphi(\mathbf{k}_{1})$ must have its own singularities (i.e., vortices) to cancel or balance those from the Berry connection, which leads to
\begin{equation}
\label{eq:integral1}
\begin{aligned}
&\lim_{\delta\rightarrow 0}\left(\int_{\mathcal{L}_{0}+\delta}-\int_{\mathcal{L}_{0}-\delta}\right)\partial_{\mathbf{k}_{1}}\varphi\cdot d\mathbf{k}_{1}=\lim_{\delta\rightarrow 0}\left(\int_{\mathcal{L}_{0}+\delta}-\int_{\mathcal{L}_{0}-\delta}\right)[\mathbf{A}_{1}(\mathbf{k}_{1})-g\mathbf{A}_{2}(\mathbf{k}_{2})]\cdot d\mathbf{k}_{1}
\\
&=\lim_{\delta\rightarrow 0}\left(\int_{\mathcal{L}_{0}+\delta}-\int_{\mathcal{L}_{0}-\delta}\right)\mathbf{A}_{1}(\mathbf{k}_{1})\cdot d\mathbf{k}_{1}-\lim_{\delta\rightarrow 0}\left(\int_{g(\mathcal{L}_{0}+\delta)}-\int_{g(\mathcal{L}_{0}-\delta)}\right)\mathbf{A}_{2}(\mathbf{k}_{2})\cdot d\mathbf{k}_{2}
\\
&=2\pi(\mathcal{C}_{1}-\mathcal{C}_{2}).
\end{aligned}
\end{equation}
Dividing Eq.~\eqref{eq:integral1} by $2\pi$ yields the net vorticity over the 2D BZ, as illustrated in Fig.~\ref{fig:chernvorticity}. For the SC pairing, $\mathcal{C}_{2}$ is the Chern number of $\ket{\xi^{(e)\star}(\mathbf{k}_2)}$ that is opposite in sign to the Chern number of $\ket{\xi^{(e)}(\mathbf{k}_2)}$. Consequently, Eq.~\eqref{eq:integral1} is just the Chern-vorticity theorem used in the main text.

It is clear that this proof highly relies on the band theory, which is no longer valid in a strongly correlated systems where band structures are not well-defined.

\section{Landau levels of surface Dirac modes of class AIII SPT}
In this section, we explain the points (i-iii) about the zeroth Landau levels of surface Dirac modes. We start by considering a minimal model of class AIII topological insulator (TI) and solving its surface Hamiltonian. For simplicity, we use the first quantization language here, under which the Hamiltonian of a minimal model of a class AIII TI can be expressed as
\begin{equation}
\label{eq:exmpmodel}
H_{\text{3DTI}}(\mathbf{k})= t \sin k_{x}\tau_{x}s_{x} \pm t \sin k_{y}\tau_{x}s_{y}+ t \sin k_{z}\tau_{x}s_{z}+(m+\sum_{i=x,y,z}\cos k_{i})\tau_{y}s_{0},
\end{equation}
which has a chiral symmetry $\tau_{z}s_{0} H_{\text{3DTI}}(\mathbf{k}) \tau_{z}s_{0}=-H_{\text{3DTI}}(\mathbf{k})$. Note that the anti-unitary chiral symmetry in second quantization language will become unitary. $\tau_{x,y,z}$ and $s_{x,y,z}$ are the Pauli matrices for orbital and spin. When $-3<m<-1$, the bulk topological invariant $\nu=+1 (-1)$ when the sign before $\sin k_{y}\tau_{x}s_{y}$ is positive (negative). Consider a surface at $z=0$, with the bulk of a 3D TI (the vacuum) residing in $z<0$ ($z>0$), then we can derive the surface localized zero energy mode at $k_{x}=k_{y}=0$, $\psi_{\text{surf},l,0}= e^{-\lambda z}u_{l}$, by solving
\begin{equation}
(-i\partial_{z}\tau_{x}s_{z}+\tilde{m}\tau_{y}s_{0})\psi_{\text{sruf},l,0}=0,
\end{equation}
where $\tilde{m}=m+3$
Since for a localized mode $\lambda=-\tilde{m}$, we can solve $u_{1}=(0,0,0,1)^T$ and $u_{2}=(1,0,0,0)^{T}$. Projecting onto the low-energy Hilbert space spanned by $\psi_{\text{surf},1,0}$ and $\psi_{\text{surf},2,0}$ (of which the Pauli matrices are denoted as $\sigma_{x,y,z}$), we can derive the Hamiltonian for surface Dirac modes shown in Eq.~(2) of the main text. Meanwhile, the chiral symmetry $\tau_{z}s_{0}$ is projected into $\sigma_{z}$ which is the $\hat{S}^{\prime}$ discussed in the main text.

When a magnetic field $B\hat{z}$ is applied, the surface Hamiltonian of the class AIII topological insulator is modified to include the effects of Landau levels. For the case where the bulk topological invariant $\nu=\pm 1$,  the surface Hamiltonian in the presence of a magnetic field $B\hat{z}$ is given by:
\begin{equation}
h_{\nu=\pm 1,B}=v\begin{pmatrix}
0 & \Pi_{x}+ i\nu \Pi_{y}
\\
\Pi_{x}- i\nu \Pi_{y} & 0
\end{pmatrix},
\end{equation}
where $\mathbf{\Pi}=\mathbf{p}+e\mathbf{A}$ is the canonical momentum including the vector potential $\mathbf{A}$, and satisfies $[\Pi_{x},\Pi_{y}]=-i e \hbar B$. In the algebraic approach for Landau level problems, it is common to define $a=(\Pi_{x}-i\operatorname{sgn}(B)\Pi_{y})/\sqrt{2e\hbar |B|}$ such that $[a,a^{\dag}]=1$, and $a\ket{0}=0$ where $\ket{0}$ represents the lowest LL with Chern number $\operatorname{sgn}(B)$. Then, for $\nu=+1$ and $\operatorname{sgn}(B)=+1$, the surface Hamiltonian is
\begin{equation}
\label{eq:hnu+1}
h_{\nu=+1,B>0}=v\begin{pmatrix}
0 & a^{\dag}
\\
a & 0
\end{pmatrix},
\end{equation}
which has its zeroth LL to be $(\ket{0},0)^{T}$ with chirality $+1$. Under the second quantization language, this means that the zeroth LL is composed of $\psi^{\dag}_{\text{surf},1,\mathbf{k}}$ that transforms as $\hat{S}^{\prime}\psi^{\dag}_{\text{surf},1,\mathbf{k}}\hat{S}^{\prime -1}=\psi_{\text{surf},1,\mathbf{k}}$. For $\nu=+1$ but $\sgn(B)=-1$, the surface Hamiltonian is 
\begin{equation}
\label{eq:hnu+1}
h_{\nu=+1,B<0}=v\begin{pmatrix}
0 & a
\\
a^{\dag} & 0
\end{pmatrix},
\end{equation}
of which the zeroth LL is  $(0,\ket{0})^{T}$ with chirality $-1$. Similarly, flipping $\nu$ will also exchange $a$ and $a^{\dag}$ and thus flip the chirality of the zeroth LL. However, whatever the chirality is, the Chern number of $\ket{0}$ is always determined by $\sgn(B)$.

\section{More details about the $\Z_{8}$ classification}
Here, we illustrate two key points: Firstly, the bosonic statistics of chiral symmetrically gapped vortices enable their condensation. Secondly, we demonstrate how the vortex proliferation argument applies specifically to the $U_s(1)\times \hat{S}$ symmetric SC phases. 

Before delving into the detailed discussion of these key points, it's important to note a distinction in the basis used. In Eqs.(3) and (5), our basis differs from that employed in Ref.~\cite{wang2014interacting} by a unitary transformation $\psi_{\text{surf},\mathbf{k}} \rightarrow U \psi_{\text{surf},\mathbf{k}}$, where 
\begin{equation}
    U=-i\exp\left(-i\frac{\pi}{4}\sigma_{x}\right).
\end{equation}
This explains why the same pairing term, $h_1$, breaks the chiral symmetry in Ref.~\cite{wang2014interacting} but not in our case.

\subsection{Statistics of chiral symmetrically gapped vortices}
For the charge order case discussed in the main text, there are eight copies of surface Dirac modes with intra-copy pairings, each of them can be described by a BdG Hamiltonian:
\begin{equation}
\label{eq:BdG1}
\begin{aligned}
H_{\text{BdG}}&=\frac{1}{2}\sum_{\mathbf{k}}\Psi_{\mathbf{k}}^{\dag}\begin{pmatrix}
k_{x}\sigma_x-k_{y}\sigma_y-\mu & -i\Delta \sigma_{y}
\\
i\Delta \sigma_{y} & k_{x}\sigma_{x}+k_{y}\sigma_{y}+\mu
\end{pmatrix}\Psi_{\mathbf{k}}
\\
&=\frac{1}{2}\sum_{\mathbf{r}}\Psi_{\mathbf{r}}^{\dag}\begin{pmatrix}
-i\partial_{x}\sigma_x+i\partial_{y}\sigma_y-\mu & -i\Delta \sigma_{y}
\\
i\Delta \sigma_{y} & -i\partial_{x}\sigma_{x}-i\partial_{y}\sigma_{y}+\mu
\end{pmatrix}\Psi_{\mathbf{r}}
\end{aligned}
\end{equation}
where $\Psi_{\mathbf{k}}\equiv (\psi_{\text{surf},1,\mathbf{k}},\psi_{\text{surf},2,\mathbf{k}},\psi^{\dag}_{\text{surf},1,-\mathbf{k}},\psi^{\dag}_{\text{surf},2,-\mathbf{k}})^{T}$ and $\Psi_{\mathbf{r}}=1/\sqrt{N}\sum_{\mathbf{k}}e^{i\mathbf{k}\cdot\mathbf{r}}\Psi_{\mathbf{k}}$ with $N$ to be the number of lattice sites in the system. Following Ref.~\cite{fu2008superconducting}, if we consider a vortex at $\mathbf{r}_{0}$ such that $\Delta=\Delta_{0}(|\mathbf{r}-\mathbf{r}_{0}|)e^{-i\theta}$ and set $\mu=0$, then there will be a localized Majorana zero mode:
\begin{equation}
\label{eq:MZM}
\gamma=\exp\left(i\frac{3\pi}{4}\right)(i\psi^{\dag}_{\text{surf},1,\mathbf{r}}+\psi_{\text{surf},1,\mathbf{r}})\exp\left(-\int_{|\mathbf{r}_{0}|}^{|\mathbf{r}-\mathbf{r}_{0}|}dr^{\prime}\Delta_{0}(|\mathbf{r}^{\prime}-\mathbf{r}_{0}|)\right).
\end{equation}
This Majorana zero mode transforms trivially under the chiral symmetry $\hat{S}^{\prime}$ as emphasized in the main text.

Since we have eight copies of Eq.~\eqref{eq:BdG1}, there will be eight Majorana zero modes trapped at one vortex, which we denote as $\gamma_{n}$ for $n=1,2,\ldots,8$. To clearly investigate the chiral transformation of states pertinent to these Majorana modes, we define $f_{i}=\gamma_{2i-1}+i \gamma_{2i}$, and define $\ket{\text{vac}}$ to be a state with all the negative-energy
levels filled in a vortex background. The chiral operator acts on $f_{i}$ and $\ket{\text{vac}}$ as
\begin{equation}
\label{eq:chiralact}
\hat{S}^{\prime}f_{i}\hat{S}^{\prime -1}=f_{i}^{\dag}, \ \hat{S}^{\prime}\ket{\text{vac}}=f_{1}^{\dag}f_{2}^{\dag}f_{3}^{\dag}f_{4}^{\dag}\ket{\text{vac}}.
\end{equation}
Then, it is straightforward to see that the fermion parity of a state will never change upon a chiral transformation. This means that the gapped, chiral symmetric ground state of a vortex, if exists, can have a well-defined fermion parity.

Have known the symmetry property, let us now demonstrate how to symmetrically gap out the eight Majorana zero modes. Following Ref.~\cite{fidkowski2010effects}, we can separate the eight Majoranas into two groups, e.g., $\gamma_{1,2,3,4}$ in the first group and $\gamma_{5,6,7,8}$ in the second group. Then, symmetric terms $\gamma_{1}\gamma_{2}\gamma_{3}\gamma_{4}$ ($\gamma_{5}\gamma_{6}\gamma_{7}\gamma_{8}$) can lead to spin-1/2 doubly degenerate ground states with a fixed fermion parity, i.e., the eigenvalue of $\gamma_{1}\gamma_{2}\gamma_{3}\gamma_{4}$ 
($\gamma_{5}\gamma_{6}\gamma_{7}\gamma_{8}$), for the first (second) group. The way to gap out the system is to introduce an chiral symmetric anti-ferromagnetic interactions between the spin-1/2 doubly degenerate ground states of the two groups, which is exactly what developed in Ref.~\cite{fidkowski2010effects}.  Importantly, the resulting gapped singlet state indeed have a well-defined fermion parity that is determined by the fermion parity of the spin-1/2 doubly degenerate ground states in the two groups. 

Given that each vortex have a well-defined fermion parity, the identical vortices are mutually local, i.e., exchanging two vortices with the same fermion parity $p_{f}$ will not lead to any extra phase for the wavefunction. This is because when exchanging two vortices, the fermions inside one vortex will cross the branch cut of the other vortex, leading to a phase $(-1)^{p_{f}}$. This phase will exactly cancel that from exchanging fermions inside the two vortices. Therefore, the statistics of the vortices are bosonic, which allows them to condensate.

\subsection{Vortex proliferation argument for the SC cases}

In the SC cases where $\mathcal{C}_{+}=\mathcal{C}_{-}=+4$,  the corresponding 3D bulk has 
$\nu_{+}=-\nu_{-}=+4$ and $\nu_{s}=8$. Consequently, the surface has eight Dirac modes -- four described by described by $h_{\nu=+1}$ and the other four described by $h_{\nu=-1}$. Since $h_{1}$ in the main text also breaks the $U_{s}(1)$ symmetry, it can gap out each copy here as in the charge order cases. In the SC cases, the important symmetry is $U_{s}(1)$. Thus, we now want to  introduce vortex associated with $U_{s}(1)$ for the pairing $\Delta$ in $h_{1}$. Mathematically, this means that in the $+$ valleys $\Delta=\Delta_{0}(|\mathbf{r}-\mathbf{r}_{0}|)e^{-i\theta}$ while in the $-$ valleys $\Delta=\Delta_{0}(|\mathbf{r}-\mathbf{r}_{0}|)e^{+i\theta}$. Consequently, the Majorana zero modes trapped in each vortex will be identical to that in the charge order cases, and thus the process to gap out and then condense the vortices discussed in last subsection can be directly applied here.

\section{Effective 1D fermion chain and finite size scaling}
The numerical evidences shown in Fig.~2 of the main text strongly suggest that no phase transition occurs as we go from the thin cylinder limit to a 2D cylinder geometry; and that $\ket{G}$ will adiabatically become a gapped, symmetric ground state with both SC order and topological order for the 2D system.
In this section we give more detailed description of the formalism and methods used for obtaining the energy gap shown in Fig.~2 of the main text. 

\subsection{Dirac Delta Potential and Effective 1D Model}
Following Refs. \cite{lee2004mott} and \cite{seidel2005incompressible}, we consider the two-body potential
\begin{equation}
H_{\text{int}} = \int d^2r d^2r' V(\mathbf{r} - \mathbf{r}') \psi_1^\dagger(\mathbf{r}) \psi_2^\dagger(\mathbf{r}') \psi_3(\mathbf{r}') \psi_4(\mathbf{r}) + \text{h.c.},
\end{equation}
where $V(\mathbf{r} - \mathbf{r}') = V_0 \delta(\mathbf{r} - \mathbf{r}')$, and $\psi_a(\mathbf{r})$ is the annihilation operator for a local fermion in the $a$-th copy/section. Subscripts $a=1,2,3,4$ respectively correspond to subscripts $(+,1), (-,1),(+,2),(-,2)$ in the main text. Projecting to the LLLs, the local fermion annihilation operators can be expressed as
\begin{equation}
\psi_a(\mathbf{r}) = \sum_n W_{an}(\mathbf{r}) c_{an},
\end{equation}
where 
\begin{equation}
W_{an}(\mathbf{r}) = \frac{1}{\sqrt{\pi^{1/2} L_y l_B}} e^{i 2 \pi n y / L_y} e^{-(x - 2 \pi n l_B^2 / L_y)^2 / 2 l_B^2}
\end{equation}
is the stripe-like wavefunction of Landau levels under the Landau gauge. And $n \in [0, N - 1]$ with
\begin{equation}
    l_B = \sqrt{\hbar / eB}
\end{equation}
being the magnetic length and
\begin{equation}
    N = L_x L_y / (2 \pi l_B^2)
\end{equation}
being the Landau degeneracy. Here, we focus on $\mathcal{C}_a = +1$ for $a = 1, 2, 3, 4$.
In the following, we will repeatedly use
\begin{equation}
\int dx dx' \partial_n^x \delta(x - x') f(x, x') = (-1)^n \int dx \partial_n^x f(x, x') \bigg|_{x' = x}.
\end{equation}
By substituting Eq. (2) into Eq. (1) and conducting the integral $\int dy dy'$, we can derive
\begin{equation}
\begin{aligned} \label{eq:Hint}
H_{\text{int}} &= \frac{V_0}{\pi L_y l_B^2} \sum_{n_1, n_2, n_3, n_4} \delta_{n_1 + n_2, n_3 + n_4} \int dx f(x) c_{1 n_1}^\dagger c_{2 n_2}^\dagger c_{3 n_3} c_{4 n_4} + \text{h.c.} \\
&= \frac{V_0}{\pi L_y l_B^2} \sum_{n_1, n_2, n_3, n_4} \delta_{n_1 + n_2, n_3 + n_4} \exp \left\{ -\frac{\alpha^2}{4} \frac{n_3^2 + n_4^2 - 2 n_1 n_2}{l_B^2 / 2} \right\} \sqrt{\pi l_B^2} c_{1 n_1}^\dagger c_{2 n_2}^\dagger c_{3 n_3} c_{4 n_4} + \text{h.c.} \\
&= \frac{V_0}{\sqrt{2 \pi} L_y l_B} \sum_{R, j, l} \exp \left\{ -\kappa^2 (l^2 + j^2) \right\} c_{1, R+j}^\dagger c_{2, R-j}^\dagger c_{3, R-l} c_{4, R+l} + \text{h.c.} \\
&= g \kappa \sum_{R, j, l} \exp \left\{ -\kappa^2 (l^2 + j^2) \right\} c_{1, R+j}^\dagger c_{2, R-j}^\dagger c_{3, R-l} c_{4, R+l} + \text{h.c.},
\end{aligned}
\end{equation}
where
\begin{equation}
    f(x) = \exp \left\{ -\frac{(x - \alpha (n_1 + n_2) / 2)^2}{l_B^2 / 2} - \frac{\alpha^2}{4} \frac{n_3^2 + n_4^2 - 2 n_1 n_2}{l_B^2 / 2} \right\}
\end{equation}
with
\begin{equation} \label{eq:defs}
    \alpha \equiv 2 \pi l_B^2 / L_y, ~~ \kappa = 2 \pi l_B / L_y,~~g = \frac{V_0}{(2 \pi)^{3/2} l_B^2}
\end{equation}
In the last step, we replace $n_1, n_2, n_3, n_4$ by $R + j, R - j, R - l, R + l$, and $R, j, l \in (-\infty, \infty)$ for systems infinite along the $x$-direction. $R, j, l$ can take half-integer values such that $R \pm j$ and $R \pm l$ are integers.

At the thin cylinder limit, $\kappa$ is large, and the leading term is just an onsite interaction
\begin{equation}
h_0 = g \kappa \sum_n c_{1n}^\dagger c_{2n}^\dagger c_{3n} c_{4n} + \text{h.c.},
\end{equation}
where $n$ takes integer values. On each site, $h_0$ lifts the degeneracy of $(c_{1n}^\dagger c_{2n}^\dagger + c_{3n}^\dagger c_{4n}^\dagger) |0\rangle$ and $(c_{1n}^\dagger c_{2n}^\dagger - c_{3n}^\dagger c_{4n}^\dagger) |0\rangle$, which have eigenenergies $\pm g \kappa$. Thus, if there is only the leading term $h_0$, i.e., in the thin cylinder limit, the system is gapped with ground state $\prod_n (c_{1n}^\dagger c_{2n}^\dagger - c_{3n}^\dagger c_{4n}^\dagger) |0\rangle$. The gap to excitation is $g \kappa$.

The question to explore next is for small $\kappa$ corresponding to the 2D cylinder geometry: will the gap close? In other words, what is the dependence of the gap on $\kappa$ and the system size? To answer this question, in the text subsection, we will use density matrix renormalization group (DMRG) to get the energy gap in systems of different sizes, and in the thermodynamic limit via a finite-size scaling. 

\subsection{Density matrix renormalization group and finite size scaling}
Since the on-site Hilbert space has four dimensions, obtaining energy gap from exact diagonalization, which, in this case, can only deal with systems up to very few sites, is not feasible. Instead we resort to DMRG which allows us to solve the energy gap in much larger systems. The DMRG computation is done using ITensor library \cite{itensor,itensor-r0.3}. 

We first investigate the behavior of the gap as a function of $\kappa$ in systems of different sizes $N$. For clarity, it is convenient to write $L_x, L_y$ in relation to $\kappa$ defined in Eq.~\eqref{eq:defs}:
\begin{equation} \label{eq:lxly}
    \frac{L_y}{l_B} = \frac{2\pi}{\kappa},~~\frac{L_x}{l_B} = \kappa N 
\end{equation}
Hence for a finite system, if $\kappa \gg 1$, we have $L_y \ll L_x$, while on the other hand if $\kappa$ is very small, $L_y \gg L_x$. These extreme cases cannot be used to faithfully approximate the 2D nature of the model. To mitigate this issue, we choose the intermediate scale $2 \ge \kappa \ge 0.5$, and calculate systems with sizes up to $N=25$ (which amounts to $100$ spin-$\frac{1}{2}$ degrees of freedoms after the Jordan-Wigner transformation.)  

We obtain the ground state using two-site DMRG with maximal bond dimensions $d \le 800$. For large $\kappa$, such as those close to $\kappa = 2$, the interaction is very localized, hence it suffices to use less bond dimensions $d\sim 300$ to have the truncation error bounded by $\epsilon \sim 10^{-9},~ \forall N \le 25$. For small $\kappa$, such as those lie within $0.5 < \kappa < 1$, the four-fermion interaction is much less localized, requiring a larger or the maximal bond dimension $d=800$; and the algorithm converges at a slightly larger truncation error $\epsilon \sim 10^{-7}$ for $N < 25$.  The excited states are thereafter calculated by energetically penalizing the lower-energy states subjected to the orthogonality constraint and the same cut-off in the bond dimensions. 
The results are as shown in Fig.~2(left) of the main text, we found that the energy spectrum remains gapped for various $\kappa$ regardless of $N$ and the aspect ratio (R). This strongly suggest that no phase transition occurs as we go from the thin cylinder limit to a 2D cylinder geometry.

Next, in order to get the 2D thermodynamic gap by the finite-size scaling, we fix the aspect ratio $R \equiv \frac{L_x}{L_y}$ while increasing the system size $N$. 
The aspect ratio can be fixed by setting
\begin{equation}
    R \equiv \frac{L_x}{L_y} \exeq \frac{\kappa^2 N}{2\pi},~~ N \in \{5,10,15,20,25\}
\end{equation}
corresponding to a set of $\kappa(N;R)$, for a each $R$, and an associated ensemble of $N$ in the set defined above:
\begin{equation} \label{eq:kapparn}
    \kappa = \sqrt{2\pi R/N}
\end{equation}
We again apply DMRG to calculate gaps $\Delta E$ in different system sizes under fixed aspect ratio $R = 1,2,3$:
\begin{align}
    \Delta E[\kappa(N;R=1)] &= \Delta E(\sqrt{2\pi /N}) ,~~ N \in \{5,10,15,20,25\}\\
    \Delta E[\kappa(N;R=2)] &= \Delta E(\sqrt{4\pi /N}) ,~~ N \in \{5,10,15,20,25\}\\
    \Delta E[\kappa(N;R=3)] &= \Delta E(\sqrt{6\pi /N}) ,~~ N \in \{5,10,15,20,25\}
\end{align}
where we apply the same upper bound for the bond dimensions, with truncation errors of the same order. 
Note that in such cases, as the number of sites $N$ increases, $\kappa$ defined in Eq.~\eqref{eq:kapparn} decreases simultaneously albeit at a different rate. It is then readily to see from Eq.~\eqref{eq:lxly} and Eq.~\eqref{eq:kapparn}, that in the limit $1/N \rightarrow 0$, we have
\begin{equation}
    N \rightarrow \infty, ~~\kappa \rightarrow 0,~~ L_y \sim O(\sqrt{N}),~~ L_x \sim O(\sqrt{N}) = O(\kappa^{-1})
\end{equation}
which gives the true 2D system in the thermodynamic limit. Hence, we can readily get the energy gap in the 2D thermodynamic limit by a finite-size scaling amenable to DMRG; the validity of the calculation can be justified by that systems of different $R$ share the same intercept, i.e. the thermodynamic gap $\Delta E \sim 0.15 g$, at $1/N \rightarrow 0$, as is shown in the Fig.~2(right) of the main text.  Consequently, $\ket{G}$ will adibatically become a gapped, symmetric ground state with SC order for the 2D system, which necessarily supports topological order according to the classification of the 3D SPT surface.

\bibliography{refs}